\documentstyle[prd,aps,amssymb]{revtex}

\def\be{\begin{equation}}
\def\fe{\end{equation}}
\def\bea{\begin{eqnarray}}
\def\fea{\end{eqnarray}}
\def\LM{{\Lambda}}
\def\X{{\!{\it X}}}
\def\Y{{\!{\it Y}}}
\def\Z{{\!{\it Z}}}
\def\sX{{\!\it \Upsilon}}
\def\sY{{\!\it\Psi}}
\def\sZ{{\!\it\Lambda}}
\def\sW{{\!\it\Theta}}
\def\A{{\!{\it A}}}
\def\B{{\!{\it B}}}

\def\a{\alpha}
\def\b{\beta}
\def\P{{\cal P}}
\def\L{{\cal L}}
\def\Lie{{\cal L}}
\def\D{{\cal D}}
\def\bg{\ominus}
\def\dbg{{\delta_\bg}}

\def\proper{\widetilde}
\def\ru{{r_\infty}}
\def\S{{\Bbb S}}
\def\V{{\Bbb V}}
\def\M{{\Bbb M}}
\def\para{\parallel}
\def\ortho{\perp} 

\def\l{\ell}
\def\D{{\cal D}}
\def\SD{ {\cal S}}
\def\N{{\cal N}}

\def\n{{\rm n}}
\def\p{{\rm p}}
\def\e{{\rm e}}
\def\s{{\rm s}}
\def\rot{{\rm rot}}
\def\vort{{\rm vort}}
\def\con{{\rm con}}
\def\cvec{\bbox}

\def\O{{\cal O}}

\def\onel{\baselineskip}

\def\O{{\cal O}}

\def\sf{{\scriptscriptstyle \text{S}}}
\def\nf{{\scriptscriptstyle \text{N}}}

\begin{document}
\draft

\wideabs{     
\title{Covariant Vortex In Superconducting--Superfluid--Normal Fluid
Mixtures \mbox{with Stiff Equation of State}} 
\author {Reinhard Prix}
\address{D\'epartement d'Astrophysique Relativiste et de Cosmologie,\\
Centre National de la Recherche Scientifique, \\Observatoire de
Paris, 92195 Meudon, France.}
\date{\today}
\maketitle 

\begin{abstract}
{\bf Abstract.} 
The integrals of motion for a cylindrically symmetric
stationary vortex are obtained in a covariant description of a mixture
of interacting superconductors, superfluids and normal fluids. The
relevant integrated stress--energy coefficients for the vortex with
respect to a vortex--free reference state are calculated in the
approximation of a ``stiff'', i.e.\ least compressible, relativistic
equation of state for the fluid mixture. 
As an illustration of the foregoing general results, we discuss
their application to some of the well known examples of ``real''
superfluid and superconducting systems that are contained as special
cases. These include Landau's two--fluid model,  uncharged binary
superfluid mixtures, rotating conventional superconductors and 
the superfluid neutron--proton--electron plasma in the outer core of
neutron stars.  
\end{abstract}
}       

\section{Introduction}
\label{secIntro}
The subject of investigation in the present work is the structure and
energy of a stationary and cylindrically symmetric quantized vortex in
an interacting multi--fluid mixture, which may consist of 
charged and uncharged superfluids and of normal fluids.
This analysis has initially been motivated by the superfluid mixture 
commonly found in neutron star models, namely in the
outer core region, where superfluid neutrons, superconducting protons
and normal electrons are generally thought to coexist.
However, due to the generality of the present approach, it is
equally well applicable to superfluid and superconducting systems
found in more common laboratory contexts, some of which will be
discussed briefly in the concluding section \ref{secDiscussion}.

The study of superfluid mixtures has a long history, beginning with
the pioneering work of Khalatnikov \cite{khal57}, later followed by
the analysis of Andreev and Bashkin \cite{ab75}, who incorporated
allowance for a (nondissipative) interaction between the
superfluids. This effect is called ``entrainment'' (sometimes also
``drag'') and plays a central role in the study of such fluid
mixtures. The model has been further extended by Vardanian and
Sedrakian \cite{vs81} to include charged fluids, and later a
Hamiltonian formulation in the Newtonian framework has been developed
by Mendell and Lindblom \cite{ml91}. 
The problem of vortices in such mixtures has been considered
especially in the context of neutron stars, namely by Sedrakian and
Shahabasian \cite{ss80}, Alpar, Langer, Sauls \cite{als84}, 
Mendell \cite{men91} and others.

The covariant vortex solution in a single
uncharged superfluid has been analyzed by Carter and Langlois
\cite{cl95a}, who have also considered the modifications due to
the compressibility of the superfluid. The present work is on the one
hand  a generalization of this analysis to arbitrary fluid mixtures,
including charged ones and their coupling to electromagnetic fields,
but on the other hand is restricted (for technical
reasons) to the case of a ``stiff'' equation of state. This ``stiff''
case is characterized by the speed(s) of sound being equal to the
speed of light, and is, within the limits of causality, the closest
analogue to the common Newtonian incompressible models. 
Compressibility effects will be subject of future work.  
Finally, we mention the previously found result \cite{cpl00} for a
Newtonian vortex in a rotating superconductor, that the (hydrodynamic)
vortex energy is strictly independent of the rotating
``normal fluid'' of positively charged ions, a result that will be
found here to hold under much more general conditions.

In the present work we will consider only stationary situations, which
has two major advantages. First, it restricts the normal fluids to be
in a state of {\em rigid} motion, and moreover in the {\em same} state
of rigid motion, because normal fluids always possess some
nonvanishing amount of viscosity and mutual friction. This even allows
to describe a solid component in the 
present framework as a ``normal fluid'', because in the rigid state
of motion the anisotropic effects of viscosity and elasticity become
irrelevant. So we can for example conveniently describe a conventional
laboratory superconductor as a superconducting--normal fluid mixture,
consisting of superconducting electrons and a ``normal'' lattice of
ions, as will briefly be discussed in the concluding section.  
The second and even more powerful consequence of stationarity is that
we can use a {\em conservative} model based on a Lagrangian formalism
that has been developed in recent years \cite{c87,cl98} in a generally
covariant language. The use of a generally covariant instead of simply
Newtonian description has also been motivated initially by the
perspective of application to neutron stars, where relativistic
effects inevitably come into play, but this approach turns out to be
generally more flexible and convenient for the hydrodynamic
description of such systems, even if relativistic effects are not
important.

The plan of this work is as follows. In Sec.~\ref{secDescription} we
introduce the relevant notions and equations of the covariant
multi--fluid formalism on which the present analysis is based.
In Sec.~\ref{secSuperfluid} we discuss the description of
superfluids in this framework and the topology of the vortex--type
configurations.  Sec.~\ref{secMongrel} introduces what we called the  
``mongrel'' representation of superfluid--normal mixtures, that
consists of choosing the {\em superfluid momenta} and the 
{\em normal currents} as the basic variables of the description, and
which will be particularly  convenient for the present problem. In
Sec.~\ref{secVortex} we specify the class of cylindrically symmetric
and stationary vortex configurations and obtain the first integrals of
motion for these solutions. 
Sec.~\ref{secReference} is devoted to the specification and
the properties of the reference state, needed to separate the
quantities attributed to the vortex from the fluid background. 
Finally, the relevant vortex stress--energy coefficients are
integrated in Sec.~\ref{secEnergy}, using the most general
hydrodynamic modelization for the vortex core, and we find that the 
``rotation energy cancellation lemma'' of \cite{cpl00} still holds
under the more general conditions of the present work.
In the concluding section \ref{secDiscussion}, we briefly illustrate
the application of the foregoing results to some of the well known 
examples of superfluid and superconducting systems.

\section{Covariant description of perfect fluid mixtures}
\label{secDescription}
The general class of (non--dissipative) mixtures of charged or
neutral perfect fluids has been shown by Carter \cite{c87} to be
describable by an elegant covariant action principle.  
In this section we will briefly
introduce the part of the formalism and notations that will be
relevant to the present work.

In the absence of electromagnetic effects, a mixture of perfect fluids 
can be described by a Lagrangian density $\LM$ that depends
only on the particle number currents $n^\a_\X$, where late Latin
indices, $\X$, $\Y$ etc., enumerate the different fluid constituents.  
Variation of $\LM$ with respect to the currents,
\be
\delta\LM = \mu_\a^\X\,\delta n^\a_\X\,,\quad
{\rm i.e.}\quad \mu^\X_\a \equiv {\partial\LM \over \partial n_\X^\a}\,,
\label{equDefMomentum}
\fe
defines the {\em dynamical} momenta per particle $\mu_\a^\X$ as the
conjugate variables of the currents $n^\a_\X$ with respect to
$\LM$. Here and in the following we use implicit summation
(except otherwise stated) over identical spacetime as well as
constituent indices.  
Legendre transformation with respect to the currents, i.e.
\be
\P \equiv \LM - n^\a_\X\,\mu_\a^\X\,,
\fe
defines the ``Hamiltonian density'' $\P$ as a function of the dynamic
momenta $\mu_\a^\X$. This function only exists for nondegenerate systems,
that is, if the functions $\mu_\a^\X(n^\b_\Y)$ defined in 
(\ref{equDefMomentum}) are invertible. The conjugate relations
can then be written as
\be
n^\a_\X = -{\partial \P \over \partial \mu_\a^\X}\,.
\label{equDefCurrent}
\fe
Furthermore, the form of these relations is constrained by the requirement
of covariance, namely $\P$ (as well as $\LM$) has to be a {\em scalar}
density, and can therefore only depend on scalars, i.e.\ on 
\mbox{$\mu^\X_\a \mu^{\Y\a}$}. 
This restricts relation (\ref{equDefCurrent}) to be of the form
\be
n^\a_\X = K_{\X\Y}\,\mu^{\Y\a}\,,
\label{equEntrainment}
\fe
where the (necessarily symmetric) matrix $K_{\X\Y}$ is defined as
\be
K_{\X\Y} \equiv - 2{\partial \P \over \partial (\mu^\X_\a \mu^{\Y\a})}\,,
\label{equDefK}
\fe
The condition of a non--degenerate system is equivalent to 
\mbox{${\rm det}\, K \not = 0$}, and so we can write the inverse relation
\be
\mu_\a^\X = K^{\X\Y}\,n_{\Y\a}\,,\quad{\rm with}\quad 
K^{\X\Y}\,K_{\Y\Z} \equiv \delta^{\,\X}_{\,\Z}\,.
\label{equEntrainment2}
\fe
In the case of noninteracting fluids, the Hamiltonian $\P$ would not
depend on crossed scalars \mbox{$\mu^\X_\a \mu^{\Y\a}$} with
\mbox{$\X\not=\Y$}, but only on diagonal terms 
\mbox{$(\mu^\X_\a \mu^{\X\a})$}. In this case the matrix $K_{\X\Y}$ 
would be diagonal, and each current would be aligned with the respective
momentum, similar to  the case of a single perfect fluid, but any
interaction terms between different fluid constituents in the
Hamiltonian will lead to nondiagonal components of $K_{\X\Y}$, and
therefore the currents will become linear combinations (in each point) 
of the respective momenta. This (nondissipative) effect is called
``entrainment'' and has first been  considered for superfluid mixtures
of $^3$He and $^4$He by Andreev and Bashkin \cite{ab75}.

Before we come to the equations of motion, we need to extend our
description to include the electromagnetic field and its
coupling to charged fluids. This is done
via  the standard ``minimal coupling'' prescription that consists of 
defining the {\em total} Lagrangian density $\L$ as
\be
\L \equiv \LM + j^\a A_\a + {1\over16\pi} F_{\a\b} F^{\b\a}\,,
\label{equDefL}
\fe
where we are using units with \mbox{$c=1$}. The electric current $j^\a$ is
defined as  
\be
j^\a \equiv e^\X n^\a_\X\,,
\label{equElCurrent}
\fe
with $e^\X$ being the charge per particle of the constituent $\X$.
The electromagnetic 2--form $F_{\a\b}$ is defined as the
exterior derivative of the gauge 1--form $A_\a$, i.e.
\be
F_{\a\b} \equiv 2 \nabla_{[\a}A_{\b]}\,,
\label{equDefF}
\fe
where square brackets indicate (averaged) index antisymmetrization.
The symbol $\nabla_\a$ denotes the usual covariant derivative,
but we note that because of the antisymmetrization, exterior
derivatives are {\em independent} of the affine connection, 
so we could as well replace $\nabla_\a$ by the partial
derivative $\partial_\a$.

The conjugate variables of the currents $n_\X^\a$ with respect to the
{\em total} Lagrangian $\L$ are the {\em canonical} momenta
$\pi^\X_\a$, defined as
\be
\pi^\X_\a \equiv {\partial \L \over \partial n_\X^\a}\,,
\fe
which can be seen from (\ref{equDefMomentum}) and (\ref{equDefL})
to be directly related to the dynamical momenta $\mu^\X_\a$, namely 
\be
\pi^\X_\a = \mu^\X_\a + e^\X A_\a\,.
\label{equPiMu}
\fe
The equations of motion are to be derived from the total Lagrangian
$\L$ via an appropriate variational principle.
Imposing invariance of the action under free (infinitesimal)
variations of the gauge field $A_\a$ leads to the Maxwell
source equation,
\be
\nabla_\b F^{\a\b} = 4\pi j^\a\,.
\label{equMaxwell}
\fe
However, the equations of motion for the fluids can not be derived via
free variations of the currents $n_\X^\a$, as this would simply lead to the 
trivial equations \mbox{$\pi^\X_\a = 0$}. This is because free variations
of the currents contain too many degrees of freedom, which results in
overdetermined equations of motion, therefore the variations have to
be {\em constrained}. 
It has been shown in \cite{c87} that variations \mbox{$\delta
n_\X^\a$} with the correct number of degrees of freedom 
are generated by infinitesimal displacements of the worldlines of 
fluid particles.   
These worldline variations satisfy the physical constraint of 
conserving the number of particles, and they result in the correct
equations of motion for the fluids. Without entering into the
technical details of this procedure (see \cite{c87,cl98}), the
resulting equation of motion for each fluid $\X$ is found as 
(no sum over $\X$)
\be
2\, n_\X^\a \nabla_{[\a}\pi^\X_{\b]} + \pi^\X_\b \,\nabla_\a n_\X^\a = 0\,,
\fe
and by contracting this equation with  $n_\X^\b$, we 
see that it implies that the currents are conserved, 
i.e.\ \mbox{$\nabla_\a n_X^\a =0$}, so the equations of motion reduce
to the simple form of a vorticity conserving flow, namely
(no sum over $\X$)
\be
n_\X^\a\, w^\X_{\a\b}  = 0\,,
\label{equEOM}
\fe
where the (canonical) vorticity 2--form $w^\X_{\a\b}$ is defined as
the exterior derivative of the canonical momentum $\pi^\X_\a$, i.e.
\be
w^\X_{\a\b} \equiv 2\nabla_{[\a}\pi^\X_{\b]}\,.
\fe
The very compact form (\ref{equEOM}) of the equation of motion can be
seen  to ``reduce''  in the nonrelativistic limit to
the (much less compact) Euler equation of a charged fluid in
electromagnetic fields, and possibly subject to further potential
forces.  This is an example that shows the advantage and convenience
of the covariant formalism, especially for more complex applications
like interacting mixtures of possibly charged fluids in
electromagnetic fields, as considered in the present analysis. 

And finally, the stress--energy tensor $T^{\a\b}$ is found \cite{cl98}
in the form
\be
{T^\a}_\b = n_\X^\a \mu^\X_\b + \P {g^\a}_\b + {1\over 4\pi}\left( 
F^{\a\lambda} F_{\b\lambda} - {1\over4}
F^{\rho\lambda}F_{\rho\lambda} \, {g^\a}_\b\right)\,,
\label{equStressEnergy}
\fe
which (in the absence of external forces) satisfies the  equation of
(pseudo) conservation, \mbox{$\nabla_\a T^{\a\b} = 0$}. 
From the form of the stress--energy tensor (\ref{equStressEnergy}) we
see that $\P$ plays the role of a generalized pressure, which reduces
to the ordinary pressure in the case of a single fluid.

\section{Properties of superfluids and topology of vortex solutions}
\label{secSuperfluid}
We want to allow for some of the fluids to be
superfluid or superconducting, and we will denote these constituents by
capital Greek indices $\sX$, $\sY$ etc.
For ``normal'' fluids (i.e.\ not superfluid or superconducting),
we will use early Latin capital indices $\A,\,\B$ etc., so a sum
over all fluids (indexed by $\X,\,\Y$ etc.) can be written as
\mbox{$\sum_\X = \sum_\A + \sum_\sX$}. 
Apart from the electric charge there seems to be no fundamental
difference between superfluids and superconductors, and therefore we
will in the following refer to them as ``uncharged'' and  
``charged superfluids'' respectively.
We note that the present treatment considers superfluids as a subclass
of perfect fluids, and will therefore represent some restrictions as
to the application to strongly anisotropic superfluid phases like they
are found in $^3$He \cite{LesHouches99}, which is governed by additional
``internal'' degrees of freedom like the spin and angular momentum  of
the Cooper pairs. But at least for situations where these additional
degrees of freedom of the order parameter can be considered as
``frozen'' and the dynamics mainly governed by the superfluid
``phase'' to be discussed in the following, the present approach
should still represent an acceptable approximation.

We distinguish the (connected) spacetime domain $\D^\sX$ occupied by
the superfluid constituent $\sX$ from the subset of its respective
``superfluid domain'' \mbox{$\SD^\sX \subseteq \D^\sX$}, which
corresponds to what is sometimes called the ``bulk''. 
In the superfluid domain $\SD^\sX$ the canonical momentum
$\pi^\sX_\a$ always obeys the constraint
\be
\pi^\sX_\a = \hbar \nabla_\a \varphi^\sX \,,
\label{equQuantisation}
\fe
where the ``phase'' $\varphi^\sX$ is a
continuously differentiable scalar on $\SD^\sX$, that can be
multi--valued, but the differences between values
in the same point are restricted to be integer multiples of $2\pi$. 
This reminds of an angle variable and reflects the role of
$\varphi^\sX$ as a quantum phase $e^{i\varphi}$.
In addition to the property of (quantized) potential flow
(\ref{equQuantisation}),  the superfluid $\sX$ in its superfluid domain
$\SD^\sX$ is {\em perfectly inviscid}. In that sense a superfluid
is probably the best representation  of a perfect fluid in nature.
On the other hand, outside its superfluid domain, i.e.\ in 
\mbox{$D^\sX \setminus \SD^\sX$}, the superfluid is not constrained to
potential flow (\ref{equQuantisation}) and can also possess some
viscosity like a ``normal'' fluid.
The property (\ref{equQuantisation}) implies that the canonical
vorticity $w^\sX_{\a\b}$ vanishes on the whole superfluid
domain $\SD^\sX$, i.e.
\be
w^\sX_{\a\b} = 2\nabla_{[\a}\pi^\sX_{\b]} = 0 \,,
\label{equIrrot}
\fe
which states that the superfluid is irrotational, and implies
that the equation of motion (\ref{equEOM}) is automatically
satisfied on $\SD^\sX$.

Irrotational flow is of course not restricted to superfluids,
and the vortex--type configurations to be discussed later have been
known long before the discovery of superfluids, 
familiar examples are tornados or water flowing out the drain of the 
bath tub. But the multi--valuedness of the ``phase'' of a perfect
fluid in a state of potential flow is {\em not} subject to a
``quantization'' condition of integer multiples of $2\pi$, and a
perfect fluid only exists as an idealization of a ``real'' fluid 
with some nonvanishing amount of viscosity, contrary to the completely
inviscid superfluids in the superfluid domain.
Furthermore there is an important energy gain associated with the
superfluid domain $\SD^\sX$, the so--called ``condensation energy''.  
Superfluids consequently try to maximize their superfluid domain
$\SD^\sX$ (and thereby to satisfy (\ref{equQuantisation})) as far as
possible in the limits of the fluid domain $\D^\sX$.

One of the most important consequences of (\ref{equQuantisation}) is
that it allows for the topologically stable flow configurations known
as vortices, which are characterized by the property that different
values  of the (multi--valued) phase $\varphi^\sX$ in the same point
can be connected by closed paths $\Gamma$ that lie entirely in the
superfluid domain $\SD^\sX$.  
As stated above, the difference can only be of the form $2\pi N^\sX$,
where the integer $N^\sX$ is called the ``winding number''. 
The winding number $N^\sX$ of a closed path 
\mbox{$\Gamma\subset \SD^\sX$} can be  written as
\be
 N^\sX = {1\over 2\pi\hbar}\oint_\Gamma \pi^\sX_\a ds^\a\,, \quad
\Gamma \subset \SD^\sX\,.
\label{equWinding}
\fe
It is evident that $N^\sX$ does not change for continuously deformed
paths  \mbox{$\Gamma\rightarrow\Gamma' \subset \SD^\sX$}, 
and $N^\sX$ is therefore a topological constant for each equivalence
class of closed paths in $\SD^\sX$. 
A nonvanishing $N^\sX$ implies that the path
\mbox{$\Gamma\subset\SD^\sX$} can not be  continuously contracted to a  
point, because it would necessarily have to cross at least one point
$P\not\in \SD^\sX$ where the phase $\varphi^\sX$ is not defined, and
therefore $\SD^\sX$ is necessarily multiply connected if there are
nonvanishing winding numbers $N^\sX$.


\section{The ``mongrel'' representation of superfluid--normal mixtures}
\label{secMongrel}
In the previous section we have seen that a superfluid on its
superfluid domain is generally characterized by a constraint
(\ref{equQuantisation}) on the (canonical) superfluid momentum,
while ``normal'' fluids are generally more easily described in terms
of their particle number currents.
For this reason it will turn out to be extremely convenient to pass
from the ``pure'' type of representation used in (\ref{equEntrainment}),
which expresses all the currents in terms of all the momenta (or
vice--versa), to a ``mongrel'' representation where the {\em
superfluid currents} and {\em normal momenta} are expressed in terms
of the {\em superfluid momenta} and {\em normal currents}. 
This type of representation has for example been used tacitly as the base of
Landau's two--fluid model for superfluid $^4$He \cite{Landau}, which
was formulated in terms of a ``superfluid velocity'', representing in
fact the irrotational superfluid momentum of (\ref{equQuantisation})
(divided by a fixed mass), and of a ``normal fluid'' velocity, which 
represents the real mean velocity of the viscous gas of
excitations. This will be seen in some more detail in the discussion
of the two--fluid model in the concluding section \ref{secDiscussion}.

In order to pass to this mongrel representation, 
we decompose the entrainment matrix $K_{\X\Y}$ into
a purely superfluid symmetric matrix $S_{\sX\sY}$, a symmetric  matrix
$V_{\A\B}$ of purely normal (``viscous'') fluids and a ``mixed''
superfluid--normal matrix $M_{\sX\A}$, so (\ref{equEntrainment}) can
be written in this decomposition as
\bea
\cvec{n}_\sX &=& S_{\sX\sY} \,\cvec{\mu}^\sY + M_{\sX\B} \,\cvec{\mu}^\B\,,\\ 
\cvec{n}_\A  &=& M_{\A\sY} \,\cvec{\mu}^\sY + V_{\A\B}\,\cvec{\mu}^\B\,, 
\label{equPure2}
\fea
where \mbox{$M_{\sX\B} = M_{\B\sX}$}. For clarity we use in this section 
{\bf bold}
typeset for denoting spacetime vectors and covectors, as the 
spacetime indices are not important here and can be put in any
consistent way.
Applying the inverse matrix $V^{-1}$ to (\ref{equPure2}), we can easily
rewrite these relations in the ``mongrel'' form
\bea
\cvec{\mu}^\A &=& -\M^\A_\sY \,\cvec{\mu}^\sY + \;\; \V^{\A\B}
\cvec{n}_\B\,, \label{equLC1}\\ 
\cvec{n}_\sX &=& \;\; \S_{\sX\sY} \;\cvec{\mu}^\sY \;\;+ \M^\B_\sX
\cvec{n}_\B \label{equLC2}\,, 
\fea
where we defined the new matrices
\bea
\V^{\A\B} &\equiv& \left(V^{-1}\right)^{\A\B} \,,\quad
\M^\A_\sY \equiv \V^{\A\B} M_{\B\sY}\,,\nonumber\\[-0.5\onel]
\label{equMatrices}\\[-0.5\onel]
\S_{\sX\sY} &\equiv& S_{\sX\sY} - M_{\sX\A} \V^{\A\B} M_{\B\sY} \,.\nonumber
\fea
In this representation it is easy to see that terms of the form 
\mbox{$\cvec{n}_\X \,\cvec{\mu}^\X$}, e.g.\ in the stress--energy tensor
(\ref{equStressEnergy}), can be written in the ``quasi separated'' form
\be
\cvec{n}_\X \cvec{\mu}^\X = 
\cvec{\mu}^\sX \,\S_{\sX\sY} \,\cvec{\mu}^\sY  + 
 \cvec{n}_\A \, \V^{\A\B} \,\cvec{n}_\B \,,
\label{equCross}
\fe
where the effect of ``mixed'' entrainment between superfluids and
normal fluids is hidden in the use of the matrix $\S$.
As we consistently wrote lower constituent indices for currents and
upper constituent indices for momenta, we can now use this convention to
introduce a very convenient and suggestive notation, namely to use
$\S_{\sX\sY}$ to {\em lower} superfluid indices $\sX,\,\sY$ etc., 
and $\V^{\A\B}$ to {\em raise} normal fluid indices $\A,\,\B$ etc.
This can formally be understood as choosing $\S$ and $\V$ as the 
{\em metric tensors} in the  respective constituent vector spaces of
the superfluids and the normal fluids, but can also just be seen as a
shorthand notation for 
\be
\cvec{\mu}_\sX \equiv \S_{\sX\sY} \,\cvec{\mu}^\sY \,,\quad\text{and}\quad
\cvec{n}^\A \equiv \V^{\A\B} \, \cvec{n}_\B \,.
\label{equMetrics}
\fe
In this notation, stress--energy contributions 
\mbox{$\cvec{n}_\X \,\cvec{\mu}^\X$} take the simple and concise form  
\be
\cvec{n}_\X \cvec{\mu}^\X = 
\cvec{n}^\A \,\cvec{n}_\A + \cvec{\mu}^\sX \,\cvec{\mu}_\sX \,,
\label{equCross2}
\fe
where all the information about entrainment has been encoded in the
respective metrics of the superfluid and normal constituent spaces. 

We note that the superfluid constraint (\ref{equQuantisation}) 
generally applies to the {\em canonical} momenta $\cvec{\pi}^\sX$, which only
in the case of uncharged superfluids coincide with the dynamical
momenta \mbox{$\cvec{\mu}^\sX = \cvec{\pi}^\sX - e^\sX
\cvec{A}$}. This implies a qualitative difference between charged and
uncharged superfluids, and it will be useful to separate the
superfluid constituent space into the two orthogonal subspaces that
are naturally defined by the superfluid ``charge vector'' with
components $e^\sX$.  
The respective subspaces are defined by parallel and orthogonal
projection via the  projection tensors
\be
\eta^\sX_\sY \equiv {e^\sX e_\sY \over (e^\sZ e_\sZ)} \,,\quad
\gamma^\sX_\sY \equiv \delta^\sX_\sY - \eta^\sX_\sY \,,
\label{equProjections}
\fe
where again we have used the notation \mbox{$e_\sX \equiv \S_{\sX\sY} e^\sY$}.
Now we can decompose constituent vectors, e.g.\ the superfluid
momenta as 
\mbox{$\cvec{\mu}^\sX = \cvec{\mu}_\para^\sX + \cvec{\mu}_\ortho^\sX$},
where 
\be
\cvec{\mu}^\sX_\para \equiv \eta^\sX_\sY \cvec{\mu}^\sY \,,\quad\text{and}\quad
\cvec{\mu}^\sX_\ortho \equiv \gamma^\sX_\sY \cvec{\mu}^\sY \,.
\fe
The subtlety of this notation is that even though a ``parallel''
constituent vector $\cvec{\mu}^\sX_\para$ only has nonvanishing
components for  charged superfluid constituents, and respectively
$\cvec{\mu}^\sX_\ortho$ only for uncharged superfluids, the {\em values}
of the respective components may depend on all the other superfluids
{\em and} normal fluids, as the projection tensors contain the
entrainment matrix $\S$.

\section{The stationary cylindrical vortex configuration}
\label{secVortex}
In this work we will consider the simplest, because maximally
symmetric type of vortex configuration, which is characterized by both
stationarity and cylindrical symmetry.
This means that there are three independent, commuting (in the sense of 
Lie brackets) symmetry generators $k^\a$, $l^\a$ and $m^\a$,
which can be taken to correspond to time translations, longitudinal
space translations (along the vortex axis) and axial rotations,
respectively. The geometric picture of the symmetry surfaces generated 
by $k^\a$, $l^\a$ and $m^\a$ are cylindrical hypersurfaces that build
a well behaved foliation of spacetime, and can therefore be
parametrized by a ``radial'' coordinate $r$.
Let us introduce the corresponding cylindrical coordinates 
\mbox{$\{x^0, x^1, x^2, x^3\} = \{t, z, \varphi, r\}$}, adapted to
these symmetries, i.e.  
\be
k^\a=\{1, 0, 0, 0\}\,,\quad l^\a=\{0, 1, 0, 0\}\,,\quad
m^\a=\{0, 0, 1, 0\}\,.
\fe
The symmetry requirements and the property of conserved currents
(\ref{equEOM}), i.e.\ \mbox{$\nabla_\a n_\X^\a=0$}, 
restrict the flow to be purely helical, i.e., to have no radial
components. Therefore the currents are 
confined to timelike hypersurfaces generated by the symmetry
vectors and can be written as
\be
n_\X^\a = \{ \, n_\X^t(r),\, n_\X^z(r),\, n_\X^\varphi(r),\, 0\, \}\,.
\label{equCurrents}
\fe
A further consequence of the symmetry is that any 
physically well defined quantity $Q$ of the flow must be
invariant under symmetry translations, which means that the
corresponding Lie derivatives must vanish, i.e.\ 
\mbox{$\Lie_\xi Q = 0$}, for $\xi^\a$ being any linear combination
(with constant coefficients) of the symmetry vectors $k^\a$, $m^\a$
and $l^\a$. 
This also holds for gauge dependent quantities like the canonical
momentum $\pi^\X_\a$, provided we fix the gauge in a way that
respects the same symmetries, i.e.\ when 
\mbox{$(\Lie_\xi A)_\a = 0$}. Such a gauge choice is given by
\be
A_\a = \{ \, A_t(r),\, A_z(r),\,  A_\varphi(r),\, 0 \,\}\,.
\label{equGaugeChoice}
\fe
The components $A_t$ and $A_z$ are still subject to the residual
gauge freedom of an additive constant, i.e.
\be
A_t \rightarrow A_t + {\cal G}_t\,,\quad A_z \rightarrow A_z + {\cal G}_z\,,
\label{equGaugeFreedom}
\fe
but because $\varphi$ is an angle variable, corresponding to a compact 
dimension, the gauge of the axial component $A_\varphi$ is completely fixed
by (\ref{equGaugeChoice}).
This is most easily seen by applying Stoke's 
theorem to a $\{r,\,\varphi\}$--surface integral over $F_{\a\b}$, 
i.e.\ \mbox{$\int d\Sigma^{\a\b} F_{\a\b} = \oint dl^\a A_\a$}, 
which in this trivial symmetric case just reduces to 
\mbox{$\int_0^{r_\infty}dr\,(dA_\varphi/dr)= A_\varphi(r_\infty)$}, 
and so  the gauge is fixed as
\be
A_\varphi(0) = 0 \,.
\label{equPhiGauge}
\fe
With the gauge choice (\ref{equGaugeChoice}), the symmetry condition
for $\pi^\X_\a$ reads  
\be
\left(\Lie_\xi \pi^\X \right)_\a = 0\,,
\label{equSymmetry}
\fe
where $\xi^\a$ can be any linear combination of the three symmetry
generators. 
The well known Cartan formula for the Lie derivative of a p--form
$w_{\a\b\gamma\ldots}$, namely
\be
(\Lie_\xi\,w)_{\a\b\gamma\ldots} = (p+1)\,\xi^\lambda \nabla_{[\lambda} w_{\a\b\gamma\ldots]} +
p\,\nabla_{[\a}\left(\xi^\lambda w_{\lambda\b\gamma\ldots]}\right)\,,
\fe
can be applied to the 1--form $\pi^\X_\a$ in
(\ref{equSymmetry}), and so we obtain the explicit symmetry condition, 
\be
2 \xi^\b \,\nabla_{[\b}\pi^\X_{\a]} + \nabla_\a \left(\xi^\b \pi^\X_\b
\right) = 0\,.
\label{equSymmetry2}
\fe
For superfluids (in the superfluid domain), the first term vanishes
because of the irrotationality property (\ref{equIrrot}), and so the
second term provides us with three independent integrals of
motion, corresponding to the three symmetry generators, namely 
\be
-E^\sX \equiv k^\a\pi^\sX_\a \,,\quad 
L^\sX \equiv l^\a\pi^\sX_\a \,,\quad 
M^\sX \equiv m^\a\pi^\sX_\a \,,
\label{equConstants}
\fe
interpretable respectively as the {\em energy},
(canonical) {\em longitudinal momentum}, and 
(canonical) {\em angular momentum} per particle.
While $E^\sX$ and $L^\sX$ are generally
subject to the residual gauge freedom (\ref{equGaugeFreedom}) of an
additive constant (except in the uncharged cases \mbox{$e^\sX=0$}),
the axial constant $M^\sX$ is {\em not}, because there is no gauge
freedom for $A_\varphi$.
In order to calculate the winding numbers $N^\sX$ of the vortex by
(\ref{equWinding}), we have to choose a path $\Gamma$ enclosing the
vortex axis. Such a path can always be continuously deformed into a
path generated by $m^\a$ alone, and so by (\ref{equConstants}) the
integration simply yields  
\be
N^\sX = {M^\sX \over \hbar}\,.
\label{equWinding2}
\fe
Therefore the constant (canonical) angular momentum per particle,
$M^\sX$, is an integer multiple of $\hbar$, the fundamental
quantum of angular momentum, and the corresponding angular momentum
``quantum number'' is just the winding number $N^\sX$.
The superfluid canonical momenta
\mbox{$\cvec{\pi}^\sX = \cvec{\mu}^\sX + e^\sX \cvec{A}$} are thereby
completely determined (in the superfluid domain) by the integrals of
motion (\ref{equConstants}) (modulo the gauge freedom
(\ref{equGaugeFreedom})), namely 
\be
\pi^\sX_\a = \{-E^\sX, L^\sX, \hbar N^\sX, 0\}\,, 
\quad \text{with}\quad N^\sX \in {\Bbb Z}\,,
\label{equSuperfluidPi}
\fe
where the vanishing of the radial component $\pi^\sX_r$ follows from
the helical direction (\ref{equCurrents}) of the currents $n_\X^\a$,
and the entrainment relation (\ref{equEntrainment}) together with
(\ref{equPiMu}) and the gauge choice (\ref{equGaugeChoice}).

In a more realistic treatment, the normal fluids are expected to
have some amount of viscosity, in which case  the condition of
stationarity, which excludes all dissipative motion, restricts all
the normal currents to be comoving with the same uniform rotation
$\Omega$, i.e. 
\be
n_\A^\a = n_\A^t \,v^\a \,,\quad\text{with}\quad 
v^\a \equiv k^\a + \Omega m^\a = \{ 1, 0, \Omega, 0\} \,.
\label{equNormalCurrent}
\fe
We could also have allowed for a constant longitudinal velocity along
$l^\a$, but this is trivially annihilated by a Lorentz boost, and so
we have chosen our reference frame at rest with respect to the longitudinal
motion of the normal fluids.
The symmetry condition (\ref{equSymmetry}) along the flowlines of the
normal fluids, i.e.\ with \mbox{$\xi^\a = v^\a$}, together with the equation
of motion (\ref{equEOM}) yields one integral of motion for each normal 
fluid, namely 
\be
-\bar{E}^\A = v^\a \pi^\A_\a \,.
\fe
With the given restrictions on the currents (\ref{equCurrents}) and
(\ref{equNormalCurrent}), the integrals of motion $E^\sX$, $L^\sX$, $N^\sX$,
$\bar{E}^\A$ and $\Omega$ are sufficient for the equations of
motion (\ref{equEOM}) to be satisfied.
But in order to actually integrate these differential equations,
one is still left with the generally nontrivial problem of
solving equations for the spacetime metric $g_{\a\b}$, together
with Maxwell's equation (\ref{equMaxwell}) for the gauge field $A_\a$.
However, for most vortex applications of practical interest (including
those in neutron stars), the gravitational self--interaction of the
vortex can be completely neglected, so the background
metric can in any case be considered as given in advance.
Furthermore, as the radial dimensions of vortices are generally much
smaller than the lengthscale of gravitational curvature, the local
spacetime metric of the vortex can safely be considered as flat,
and so in cylindrical coordinates we can write it as
\be
ds^2 \equiv g_{\a\b}\,dx^\a \,dx^\b = -dt^2 + dz^2 + r^2\,d\varphi^2 + dr^2 \,.
\label{equMetric}
\fe
The remaining differential equation to be solved is (\ref{equMaxwell}) 
for the electromagnetic gauge field $A_\a$. The necessary coefficients
of the metric connection can easily be calculated for the flat metric
(\ref{equMetric}), and we find the explicit Maxwell equations for the
gauge field $A_\a$ in the form
\be
\left( r A_t'\right)' = 4\pi r j^t \,, \quad
-\left(r A_z'\right)' = 4\pi r j^z \,,\label{equMax1}
\fe
\be
-\left({A_\varphi' \over r}\right)' = 4\pi r j^\varphi\,, \label{equMaxwell2}
\fe
where the prime denotes differentiation with respect to $r$.
Equations (\ref{equMax1}) describe a radial
electric field $A_t'$ created by the charge distribution $j^t$,
and an axial magnetic field $A_z'$ around a longitudinal current
$j^z$. 
These equations will result in exponentially ``screened'' solutions,
typical of charged superfluids. As we saw in section
\ref{secSuperfluid}, the vortex is characterized by nonvanishing
winding numbers $N^\sX$, which by (\ref{equPiMu}) and
(\ref{equEntrainment}) are seen to be directly related to the axial components
$j^\varphi$ and will result in a screened longitudinal magnetic field 
$B_z$, which is conventionally defined as
\be
B_z \equiv {A'_\varphi \over r} \,.
\label{equDefB}
\fe

\section{Reference state and vortex properties}
\label{secReference}

\subsection{The reference state}
In the previous section we have completely specified the fluid
configuration containing a vortex, but in order to separate
the quantities attributed to the vortex from the fluid
``background'', we first have to specify this reference ``background''
state, which will be denoted by the subscript $\bg$. 
For any quantity $Q$, the part \mbox{$\dbg Q$} attributed to the vortex is
defined as the difference with respect to the corresponding reference
value $Q_\bg$, i.e. 
\be
\dbg Q \equiv Q - Q_\bg \,.
\fe
The reference state should respect at least the same symmetries as the
vortex state, and can therefore, by the reasoning in
Sec.~\ref{secVortex}, be characterized completely by constants
$E^\sX_\bg$, $L^\sX_\bg$, $N^\sX_\bg$, $\bar{E}^\A_\bg$ and
$\Omega_\bg$. 
Furthermore, we naturally want the the reference background to be 
``vortex free'', which means that the topological constants
characterizing a vortex have to vanish, i.e.\ \mbox{$N^\sX_\bg = 0$}. 
Another natural prescription is that the uniform rotation of the
normal fluids should be the same in the reference state as in the
vortex state, i.e.\ \mbox{$\Omega_\bg = \Omega$}.
However, there is no such ``natural'' choice for the remaining
constants $E^\sX_\bg$, $L^\sX_\bg$ and $\bar{E}^\A_\bg$, if one allows
for compressibility of the fluids. The compressibility is described by
the fact that the entrainment matrix (\ref{equDefK}) is in general a
function of the momentum scalars,  i.e.\ 
\mbox{$K_{\X\Y} = K_{\X\Y}(\mu^V_\a \mu^{W\a})$}, and 
therefore, if  \mbox{$(\mu^V_\a \mu^{W\a})_\bg \not=\mu^V_\a \mu^{W\a}$},
this generally entails that \mbox{$K_{\X\Y} \not = K^\bg_{\X\Y}$}.
Now, if we consider for example the $t$ component of the relation
(\ref{equEntrainment}) between currents and momenta, and if for
illustration we suppose for a moment that there are no normal
fluids, then \mbox{$n^t_\sX = K_{\sX\sY}\mu^{\sY t}$}, and 
\mbox{$n^t_{\sX\bg} = K^\bg_{\sX\sY}\mu_\bg^{\sY t}$}.
Choosing for example the straightforward reference constants
\mbox{$E^\sX_\bg = E^\sX$} and \mbox{$L^\sX_\bg = L^\sX$}, leads to
changed particle densities \mbox{$n^t_{\sX\bg} \not= n^t_\sX$}, and
especially changed {\em mean} particle number densities (in the region
of integration with the upper cutoff radius $\ru$),
i.e.\ \mbox{$\overline{n^t_{\sX\bg}} \not= \overline{n^t_\sX}$}.  
We see that with this choice of reference constants, we compare a vortex
state with a reference state that does not have the same number of
particles in the region of integration. 
Another physically interesting choice of reference state would
therefore rather consist in readjusting the reference constants
$E^\sX_\bg$ in such a way as to obtain the same {\em mean} particle
number densities (and therefore total number of particles in the
region of integration) in the reference state. These different choices
have been analyzed and properly accounted for in \cite{cl95b} for the
case of a vortex in an uncharged superfluid, and are found to be 
inequivalent to each other, even in the limit \mbox{$\ru\rightarrow \infty$}.

Due to the additional complications of multiple entrainment
and charged fluids in the present analysis, we will postpone this problem
of compressibility effects to future work, and restrict our attention
here to the simpler case of a ``stiff'' equation of state that
is characterized by a constant entrainment matrix, i.e.
\be
{\partial K_{\X\Y} \over \partial(\mu^V_\a \mu^{W\a})} = 0 \quad
\Longrightarrow \quad K_{\X\Y}^\bg = K_{\X\Y}\,.
\label{equStiff}
\fe
In this ``stiff'' case, the most natural reference state is unambiguously 
characterized just by choosing the longitudinal superfluid momentum
components $E^\sX_\bg$, $L^\sX_\bg$ to be the same as in the vortex
state, i.e. 
\be
\pi^{\sX\bg}_\a \equiv \{-E^\sX, L^\sX, 0, 0\,\} \,,
\label{equRefSuperfluid}
\fe
while the constants $\bar{E}^\A$ can be fixed by taking the normal
particle densities to be unchanged with respect to the vortex state,
i.e. 
\be
n_{\A\bg}^\a \equiv n_\A^t v^\a \,, \quad
\text{where}\quad v^\a = \{1, 0, \Omega, 0\}\,.
\label{equRefNormal}
\fe
Due to the assumption of a stiff equation of state (\ref{equStiff}),
all longitudinal current components $n^t_\X$ and $n^z_\X$ remain 
unchanged in the reference state.
Furthermore we will assume the electric current to vanish in the
reference state, i.e. 
\be
j^\a_\bg = 0\,,
\label{equRefElectric}
\fe
which implies that the longitudinal electric current
also vanishes in the vortex state, 
\be
j^t = j^t_\bg =0\,,\quad\text{and}\quad
j^z = j^z_\bg =0\,,
\fe
and so we also have from (\ref{equMax1}) (in an appropriate gauge) 
\be
A_t = A_t^\bg = 0\,,\quad\text{and}\quad
A_z = A_z^\bg = 0\,.
\label{equRefA}
\fe
The reference state is now completely fixed by the properties
(\ref{equRefSuperfluid}), (\ref{equRefNormal}) and
(\ref{equRefElectric}). The vortex modifies only the
$\varphi$ components of currents and momenta, so it will be convenient
to introduce for covectors $Q_\a$ the short notation  
\mbox{$\proper{Q} \equiv \dbg Q_\varphi$} for the part of the
$Q_\varphi$ that is due to the vortex, and  
\mbox{$Q_\bg \equiv Q^\bg_\varphi$} for the part that is still present
in the reference state, e.g.
\be
\mu^\sX_\varphi = \proper{\mu}^\sX + \mu_\bg^\sX \,,\quad\text{and}\quad
A_\varphi = \proper{A} + A_\bg \,.
\fe
From (\ref{equSuperfluidPi}) and (\ref{equRefSuperfluid}) it is easy
to see that 
\be
\proper{\mu}^\sX = \hbar N^\sX - e^\sX \proper{A} \,,\quad\text{and}\quad 
\mu^\sX_\bg = -e^\sX A_\bg \,.
\label{equMus}
\fe

\subsection{The London field}
Contrary to the longitudinal components $A^\bg_t$ and $A^\bg_z$, the
axial gauge field $A_\bg$ in the reference state will not be trivial, 
due to the uniform rotation of the charged normal fluids.
The Maxwell equation (\ref{equMaxwell2}) for the $\varphi$
component in the reference state, i.e.\ \mbox{$(A'_\bg/r)'=0$}, allows
for a uniform magnetic field $B_\bg$ in $z$ direction (defined as in
(\ref{equDefB})), namely by integration and using (\ref{equPhiGauge})
one gets,  
\be
B_\bg \equiv {A'_\bg \over r} = {2\over r^2} A_\bg=\text{const.} \,,
\label{equLondon}
\fe
where $B_\bg$ is in fact the well known uniform London field of 
rotating superconductors.
An explicit expression for the London gauge field $A_\bg$ can be
obtained simply from the reference property 
\mbox{$j_\bg^\varphi = e^\X n_{\X\bg}^\varphi =0$}, together with the
``mongrel'' entrainment expression (\ref{equLC2}), and relation
(\ref{equMus}), which yields 
\be
A_\bg = 
r^2  \,(e^\sY e_\sY)^{-1} \left(e^\A + e^\sX \M_\sX^\A\right)
n_\A^\varphi \,,
\fe
and after using (\ref{equNormalCurrent}) to write
\mbox{$n_\A^\varphi = \Omega \, n_\A^t$}, we get the London field
$B_\bg$ as  
\be
B_\bg = 2\Omega \,
(e^\sY e_\sY)^{-1} \left(e^\A + e^\sX \M_\sX^\A\right) n^t_\A \,.
\label{equLondon1}
\fe
The London field $B_\bg$ is seen to be proportional to the uniform
rotation $\Omega$ of the normal fluids. 
If we now use the additional property of the vanishing charge density
(\ref{equRefElectric}) in the reference state, i.e.\ \mbox{$j_\bg^t=0$}, 
then we can finally obtain the very simple expression for the London
field, 
\be
B_\bg = -2\Omega \,
{e^\sX \S_{\sX\sY} E^\sY \over e^\sZ \S_{\sZ\sW} e^\sW}  = 
-2\Omega {e_\sX E^\sX \over e_\sY e^\sY} \,,
\label{equLondon2}
\fe
where we have used the notation of lowering
and raising constituent indices via the matrix $\S$ introduced in
Sec.~\ref{secMongrel}.
If we consider in particular the case of a single charged superfluid
with mass per particle $m$ and charge per particle $e$, this
expression in the Newtonian limit, where \mbox{$E^\sX \rightarrow m$},
reduces  to the well known expression 
\mbox{$B_\bg = -2\Omega m / e$}. The question of whether $m$ in this
formula should represent the bare mass or some ``effective'' mass per
particle will be discussed briefly in the concluding
section~\ref{secDiscussion}.

\subsection{The Magnetic field of the vortex}
The reference state properties (\ref{equRefSuperfluid}) and
(\ref{equRefNormal}) further allow us to rewrite the axial
current $j^\varphi$ in the form 
\mbox{$j^\varphi = e^\sX (n_\sX^\varphi - n_{\sX\bg}^\varphi)$}, and
with (\ref{equLC2}) we obtain the compact form
\be
j^\varphi = 
{1\over r^2} e^\sX \,\S_{\sX\sY}\,\proper{\mu}^\sY = 
{1\over r^2} e_\sX \, \proper{\mu}^\sX
\,.
\fe
Inserting this into the  corresponding Maxwell equation
(\ref{equMaxwell2}) gives 
\be
e^\sX \, \S_{\sX\sY} \, \proper{\mu}^\sY = 
-{r\over4\pi} \proper{B}'\,,
\label{equBessel1}
\fe
which can be written more explicitly as a differential equation for
$\proper{A}$, containing the winding numbers $N^\sX$ as parameters,
namely
\be
(e^\sY e_\sY) \proper{A} = \hbar \,(e_\sY N^\sY) + 
{r\over 4\pi} \proper{B}' \,,
\label{equBessel2}
\fe
where the longitudinal magnetic field of the vortex, 
\mbox{$\proper{B}=\dbg B_z$}, is defined following (\ref{equDefB}) as  
\mbox{$\proper{B}(r) \equiv {\proper{A}'(r)/ r}$}.
This second order differential equation for $\proper{A}$ (or
$\proper{B}$) is of the modified Bessel type, and the asymptotic behavior
of the solutions in the limit \mbox{$r\rightarrow \infty$} can be
derived directly from this equation, namely (where ``$\sim$'' means
asymptotically proportional) 
\bea
\proper{B} \sim \proper{B}' \sim  e^{-r/\l}\,, \nonumber\\[-0.5\onel]
\label{equAsymptInfty}\\[-0.5\onel]
\lim_{r\rightarrow \infty} \proper{A} = 
\hbar\, {e_\sY N^\sY \over e^\sX e_\sX}\,,\nonumber
\fea
where $\l$ is the so--called London penetration depth, which
is given by the expression
\be
\l^{-2} \equiv 4\pi \, e^\sY e_\sY \,.
\label{equPenetration}
\fe
In the Newtonian limit of a single superfluid with charge per particle 
$e$, mass per particle $m$ and a particle number density $n$, the
matrix $\S$ reduces to $n/m$, and (\ref{equPenetration}) reduces
to the standard expression \mbox{$\l^{-2} = 4\pi e^2 n /m$}.

The total electromagnetic flux of the vortex, 
\mbox{$\Phi\equiv \oint \proper{A}_\a dx^\a$}, for a circuit at
sufficiently large radial distance, is easily seen from
(\ref{equAsymptInfty}) to be given as
\be
\Phi = 2\pi \hbar \,{e_\sY N^\sY \over e^\sX e_\sX}\,,
\label{equFlux}
\fe
which again reduces to the standard expression 
\mbox{$\Phi=N (2\pi\hbar/e)$} in the Newtonian limit of a single charged
superfluid with charge per particle $e$.
The explicit solution of equation (\ref{equBessel2}) is expressible in
terms of the (modified) Bessel functions $K_0$ and $K_1$, namely
\bea
\proper{B}(r) &=& C_0 \,K_0(r/\l) \,,\nonumber\\[-0.5\onel]
\label{equSolution}\\[-0.5\onel]
\proper{A}(r) &=& {\Phi\over 2\pi} - C_0 \, {r \l} \, K_1(r/\l) \,.\nonumber
\fea
This solution is only valid in the ``common superfluid domain'',
i.e.\ in \mbox{$\bigcap_\sX \SD^\sX$}, where all the constant winding numbers
$N^\sX$ are defined. From the divergence of $\proper{B}(r)$ on the
axis it is evident that the common superfluid domain must have a
finite separation, $\xi$ say, from the axis, which can be used to
define what is usually called the ``vortex core'', with $\xi$ being
the ``core radius''. 
The constant of integration $C_0$ is to be determined from
the matching of (\ref{equSolution}) with the ``inner'' vortex
solution, i.e.\ for \mbox{$r\le \xi$}.
By integrating (\ref{equSolution}) for \mbox{$r\ge \xi$}, we 
get  the vortex flux outside the core, 
i.e.\ \mbox{$\Phi - \Phi_{\rm core}$}, and so $C_0$ can be expressed
in terms of the quantities $\xi$ and the core magnetic flux
$\Phi_{\rm core}$, namely 
\be
C_0 = {\Phi - \Phi_{\rm core} \over 2\pi \l^2 x_0 K_1(x_0) }\,,
\label{equC0}
\fe
where $x_0$ is the rescaled core radius, \mbox{$x_0 \equiv \xi / \l$},
which corresponds to the inverse of the Ginzburg--Landau parameter
\mbox{$\kappa \equiv \l/\xi$} of the Ginzburg--Landau model. The limit of an
extreme type--II superconductor is characterized by 
\mbox{$\kappa \rightarrow \infty$}, i.e.\ \mbox{$x_0\rightarrow 0$}, 
\mbox{$x_0 K_1(x_0) \rightarrow 1$}, so the core structure becomes
negligible, \mbox{$\Phi_{\rm core} \ll \Phi$}, and we get
\be
C_0 \simeq {\Phi \over 2\pi\l^2} = 4\pi \hbar \, e_\sY N^\sY
\,,\quad\text{for}\quad  \l \gg \xi\,.  
\fe

\section{The Vortex energy}
\label{secEnergy}
In this section we will consider the ``macroscopic'' properties of the
vortex, namely its total energy per unit length and the tension of the
vortex line. These quantities are obtained by integrating the local
stress--energy tensor of the vortex, \mbox{$\dbg {T^\a}_\b$},
over the spatial section \mbox{$\{r,\varphi\}$} orthogonal to the
(``longitudinal'') vortex symmetry axes, whose coordinates are the
subset \mbox{$\{x^i\} = \{t, z\}$}, for \mbox{$\{i\}=\{0, 1\}$}. 
The local stress--energy coefficients  of the vortex are
seen from (\ref{equStressEnergy}) to have the form
\bea
\dbg {T^\a}_\b &=& \dbg\left( n_\X^\a \mu^\X_\b\right) + 
{1\over4\pi} \dbg\left( F^{\a\lambda} F_{\b\lambda}\right) \nonumber\\[0.3\onel]
& &\hspace*{0.8cm}+ \left[ \dbg \P - 
{1\over16\pi}\dbg\left( F^{\rho\lambda} F_{\rho\lambda}\right)\right] 
{g^\a}_\b \,.
\label{equRecall}
\fea
The ``sectional'' \mbox{$\{r,\varphi\}$}--integral is only meaningful
for quantities that are scalars with respect to the sectional
coordinates $r$ and $\varphi$, and so we have to consider only the 
``longitudinally'' projected tensor $\dbg {T^i}_j$.
Another ``sectional'' scalar of the stress--energy tensor is the
trace of the orthogonally projected components, which defines the
local lateral pressure $\Pi$ of the vortex, 
\be
2 \Pi \equiv \dbg\left({T^\a}_\a - {T^i}_i\right) \,. 
\label{equLateral}
\fe
In the case of a ``stiff'' equation of state (\ref{equStiff}), the
Taylor expansion of $\P(\mu^\X_\a\mu^{\Y\a})$ around the reference
state value \mbox{$\P_\bg \equiv \P((\mu^\X_\a\mu^{\Y\a})_\bg)$} 
has only two terms (using (\ref{equDefK})), namely 
\be
\P(\mu^\X_\a\mu^{\Y\a}) = \P_\bg - 
{1\over 2} K_{\X\Y} \dbg(\mu^\X_\a\mu^{\Y\a})\,.
\fe
The mongrel representation (Sec.~\ref{secMongrel}) is particularly
convenient to evaluate contributions of this type, because by the
reference property (\ref{equRefNormal}) we have 
\mbox{$\dbg(\cvec{n}_\A \V^{\A\B} \cvec{n}_\B)=0$}, and so we find,
using (\ref{equEntrainment}) and (\ref{equCross}), 
\be
K_{\X\Y} \dbg(\cvec{\mu}^\X \cvec{\mu}^\Y) = 
\dbg(\cvec{n}_\X \cvec{\mu}^\X) = 
\S_{\sX\sY} \dbg( \cvec{\mu}^\sX \cvec{\mu}^\sY )\,.
\fe
The relevant contributions (\ref{equRecall}) of  \mbox{$\dbg
{T^\a}_\b$} are now straightforward to evaluate, and are found to be
given by 
\be
\dbg\left( n_\X^i \mu^\X_j \right) = 0\,,\quad
\dbg\left(F^{i\lambda}F_{j\lambda}\right) = 0\,,
\fe
\be
\dbg\left(F^{\a\b}  F_{\a\b}\right) = 2 \proper{B}^2 + 4 \proper{B}B_\bg \,,
\fe
\be
\dbg {T^\a}_\a = -\dbg\left(n_\X^\a \mu^\X_\a \right) = 2\dbg\P\,,
\fe
\be
\dbg\P = - {1\over 2r^2} \S_{\sX\sY} \, \left( 
\proper{\mu}^\sX \, \proper{\mu}^\sY + 
2 \proper{\mu}^\sX \, \mu_\bg^\sY \right)\,.
\fe
Putting these results into the expression for the vortex
stress--energy tensor (\ref{equRecall}), we find that the
longitudinally projected tensor \mbox{$\dbg {T^i}_j$} is proportional 
to the unit tensor, i.e  
\be
\dbg {T^i}_j = - \proper{T} {g^i}_j \,,
\label{equ69}
\fe
with
\be
\proper{T} = {1\over 16\pi} \dbg\left(F^{\a\b}F_{\a\b}\right) - \dbg \P\,,
\fe
and so the vortex energy density, $\dbg T^{00}$, is equal
to the (local) longitudinal tension of the vortex, $-\dbg T^{zz}$, a
property that is characteristic of the stiff equation of state
(\ref{equStiff}). 
The vortex energy per unit length $U$ is defined as the sectional
integral 
\be
U \equiv - 2\pi\int_0^\ru dr\,r\,\dbg {T^0}_0 = 2\pi \int_0^\ru
dr\,r\,\proper{T} \,.
\label{equVortexEnergy}
\fe
The energy density $\proper{T}$ can be decomposed into two parts, 
\be
\proper{T} = \proper{T}_\vort + \proper{T}_\rot,
\label{equ70}
\fe
where $\proper{T}_\vort$ is the part that is independent of the
rotation $\Omega$ of the normal fluids,
\be
\proper{T}_\vort =
{1\over2r^2} \proper{\mu}_\sY \proper{\mu}^\sY 
+ {1\over 8\pi} \proper{B}^2 \,,
\label{equTvort}
\fe
while $\proper{T}_\rot$ is proportional to $\Omega$ (via $B_\bg$, see
equ.~(\ref{equLondon2})), 
\be
\proper{T}_\rot = B_\bg \left(
{1\over 4\pi}\proper{B} - {1\over 2} e_\sY \proper{\mu}^\sY \right) \,,
\fe
and the lateral pressure $\Pi$, defined in (\ref{equLateral}),
is found to be given by
\be
\Pi = {1\over8\pi}\left[ \proper{B}^2 + 
2 \proper{B} B_\bg \right] \,.
\fe
Expression (\ref{equTvort}) for $\proper{T}_\vort$ can be transformed
using Maxwell's equation (\ref{equBessel1}) into the ``nearly
integrated'' form 
\bea
\proper{T}_\vort &=& {\hbar^2\over 2 r^2} 
\left[ N^\sY N_\sY - {(e_\sX N^\sX)^2\over e^\sY e_\sY}\right]
\nonumber\\[0.3\onel]
& & \hspace*{0.8cm} -  {e_\sX N^\sX \over e_\sY e^\sY} 
{\hbar\over 8\pi r} \proper{B}' 
+ {1\over 8\pi r}\left( \proper{A} \proper{B} \right)' \,.
\label{equTV}
\fea
The easiest way to see this is to first expand only one
$\proper{\mu}^\sX$ in (\ref{equTvort}) using (\ref{equMus}) and
apply (\ref{equBessel1}), then expand the remaining $\proper{\mu}^\sY$
and use the second form of Maxwell's equation (\ref{equBessel2})
for $\proper{A}$.  In order to regroup the derivatives, one also has
to expand one $\proper{B}$ as $\proper{A}'/r$ in the last term of
(\ref{equTvort}). 
In a similar way, $\proper{T}_\rot$ can be reduced to
\be
\proper{T}_\rot = {B_\bg \over 8\pi r} \left( r^2\proper{B} \right)' \,.
\label{equTrot}
\fe 
As anticipated from the divergence of the magnetic field
(\ref{equSolution}) on the vortex axis, we encounter the same problem in
the energy density (\ref{equTV}). This well known fact is 
due to the constant superfluid (canonical) angular momentum per
particle, \mbox{$\pi^\sX_\varphi = \hbar N^\sX$}, in the superfluid
domain $\SD^\sX$. Therefore each superfluid with a nonvanishing
winding number \mbox{$N^\sX\not=0$}, must have some finite ``core''
region separating the respective superfluid domain $\SD^\sX$ from the
vortex axis. The actual size of the respective
core region is determined by a trade--off between the loss of
condensation energy associated with the core region, and the diverging
energy density (\ref{equTV}) in the superfluid domain.
The detailed description of this superfluid--normal transition would
ask for either a microscopic theory, or at least some
phenomenological, e.g.\ Ginzburg--Landau type description of the involved
superfluids. However, such detailed descriptions turn out to be
unnecessary for our 
present purpose, as we can proceed on the basis of a very general
hydrodynamic description of the vortex core, based only on the
necessary ``minimal assumptions'' needed to avoid the energy divergence. 
Namely, as the superfluid constraint (\ref{equQuantisation}) does no
longer apply in the respective ``core'' regions, the (canonical)
angular momentum  $\pi^\sX_\varphi$ there is not quantized, and
is allowed to depend on the radial variable $r$. The winding number
$N^\sX$ is strictly speaking not defined in the core region, but we can keep
the same symbol as a shorthand notation for \mbox{$\pi^\sX_\varphi /\hbar$},
so we cast our general description of the core region in the simple form
\be
N^\sX(r) = \left\{ \begin{array}{ll}
		N^\sX \;\in \; {\Bbb Z} \quad&\text{for}\quad r > \xi \\
		\N^{\,\sX}(r) &\text{for}\quad r\le \xi
	\end{array} \right. \,,
\label{equCore}
\fe
where $\N^{\,\sX}(r)$ is a continuous, monotonic function, which
has to ensure the vortex energy density $\proper{T}$ to remain finite
on the vortex axis, i.e.\ in the limit \mbox{$r \rightarrow 0$}.
Note that the ``core radius'' $\xi$ is defined, as in
Sec.~\ref{secVortex}, as the radial distance of the ``common superfluid
domain'' \mbox{$\bigcap_\sX \SD^\sX$} for the vortex axis, and is
therefore the maximum core radius of the individual superfluids. This
obviously does not restrict 
the generality of the core description (\ref{equCore}), as the
$\N^\sX(r)$ are allowed to remain constant until some smaller radius
\mbox{$\xi^\sX < \xi$}.
In order to have a regular behavior of the energy density
$\proper{T}$ near the axis, it is sufficient to demand that $\N(r)$
and $\proper{A}(r)$ vanish on the vortex axis {\em at least} as  
\be
\N(r) \sim r \,,\quad\text{and}\quad
\proper{A} \sim r^2 \quad\text{for}\quad r\rightarrow 0 \,,
\label{equAsymptAxis}
\fe
where by ``$\sim$'' we mean ``asymptotically proportional'' (and not
necessarily equal).
This phenomenological description is based on only two parameters, the
``core radius'' $\xi$ and the core condensation energy per unit length
$U_\con$.  These two phenomenological parameters would have to be
determined either from experiment or from a microscopic theory, 
but the model is now sufficiently determined to allow the integration
of the vortex energy, without the need  of further assumptions
concerning the underlying physical processes of superfluidity. 

The total vortex energy per unit length is
\be
U = U_\con + U_\vort + U_\rot \,,
\label{equTotalEnergy}
\fe
where according to (\ref{equVortexEnergy}) and (\ref{equ70}) we have defined
\bea
U_\vort &\equiv& 2\pi\int_0^\ru dr\,r\,\proper{T}_\vort\,,
\nonumber\\[-0.5\baselineskip] 
\label{equDefUs}\\[-0.5\baselineskip]
U_\rot &\equiv& 2\pi\int_0^\ru dr\,r\,\proper{T}_\rot \,.\nonumber
\fea
The energy contribution $U_\rot$, which is proportional to the
rotation $\Omega$ of the normal fluids, is found from (\ref{equTrot})
to be
\be
U_\rot = 
{B_\bg\over 4} \left. \left( r^2 \proper{B} \right) \right|_0^\ru = 
0 \,,
\label{equRotCancel}
\fe
where the vanishing of the integral follows from the asymptotic
properties (\ref{equAsymptInfty}) and (\ref{equAsymptAxis}) of the 
magnetic field $\proper{B}$. 
In the Newtonian description of a rotating superconductor, 
the vortex energy was already found \cite{cpl00} to be unchanged by
the rotating charged background, and this lemma is seen here to still
hold under quite general conditions: \vspace{1em}\\
{\bf Rotation energy cancellation lemma:} {\em 
The ``hydrodynamic'' energy per unit length (i.e.\ excluding the core
condensation energy $U_\con$) of a cylindrically symmetric
and stationary vortex in a ``stiff''  mixture of
interacting superfluids, superconductors and normal fluids
(\ref{equStiff}) is independent of the uniform rotation rate $\Omega$
of the normal fluids, despite the fact that the radial distribution of
the  hydrodynamic energy density is modified by $\Omega$, as seen in 
(\ref{equTrot}).}
\vspace{1em}\\
The vortex energy contribution $U_\vort$ in  (\ref{equDefUs}) is
found by integrating (\ref{equTV}), which yields
\bea
U_\vort &=& \pi \hbar^2 \left[ 
N^\sY N_\sY - { (e_\sX N^\sX )^2\over e_\sY e^\sY}\right] 
\text{ln}{\ru\over\zeta} + {\Phi \proper{B}(\eta) \over 8\pi}\,,
\nonumber\\[0.5em]
& &\quad\text{with}\quad
0 < \zeta,\, \eta  \le \xi \,, \label{equResult1}
\fea
where we used the asymptotic properties (\ref{equAsymptInfty}), 
(\ref{equAsymptAxis}), and the 
(first) mean value theorem of integration with the
intermediate values $\zeta$ and $\eta$, after a partial integration in
the core region.
We recognize two qualitatively different energy contributions; the
first one from a  ``global'' vortex, diverging logarithmically with
the upper cutoff radius  $\ru$, which is characteristic for vortices
in uncharged superfluids, and the second one from a ``local'' vortex,
whose energy contribution has the standard ``axis field'' form 
\mbox{$\Phi B(0)/8\pi$}, which is typical for vortices in charged
superfluids. 

Using the decomposition into charged and uncharged superfluid
subspaces via the charge projection tensors defined in
(\ref{equProjections}), we can rewrite the first term in brackets in
the form 
\be
\left[ N^\sY N_\sY - {(e_\sX N^\sX)^2 \over (e^\sY e_\sY)} \right] = 
N^\sX_\ortho N_\sX^\ortho \,.
\fe
Concerning the second term in (\ref{equResult1}), if
the magnetic field $\proper{B}(r)$ is slowly varying inside the vortex 
core, then we can approximately replace 
\mbox{$\proper{B}(\eta) \approx\proper{B}(\xi)$}, and use the explicit
expression (\ref{equSolution}) with (\ref{equC0}) and (\ref{equFlux})
to write 
\be
\proper{B}(\xi) = 4\pi\hbar \,(e_\sY N^\sY_\para) \left(
1 - {\Phi_{\rm core}  \over \Phi} \right) {K_0(x_0) \over x_0
K_1(x_0)} \,,
\label{equVortexB}
\fe
where \mbox{$x_0 \equiv \xi / \l$}.
In the extreme type--II limit, where the core structure becomes
negligible, i.e.\ in the limit  \mbox{$\kappa = 1/x_0 \gg 1$}, where
\mbox{$\Phi_{\rm core} \ll \Phi$},  
\mbox{$K_0(\xi/\l) \approx {\rm ln}(\l/\xi)$}, and  
\mbox{$x_0 K_1(x_0)\approx 1$}, equation (\ref{equResult1}) with
(\ref{equVortexB}) finally gives the simple expression for the vortex
energy 
\be
U_\vort = \pi \hbar^2\, \S_{\sX\sY}\,\left[ 
(N^\sX_\ortho N^\sY_\ortho) \,\text{ln} {\ru\over \xi} +
(N^\sX_\para N^\sY_\para)\,\text{ln} {\l \over \xi}
\right] \,. 
\label{equResultII}
\fe
This ``quasi separated'' form clearly shows the respective
contributions from a global vortex and a local vortex,
but as mentioned above, even for vortices which have 
nonvanishing winding numbers only in either charged or uncharged
constituents, there will generally be contributions from {\em both} terms,
due to the entrainment matrix $\S$ involved in the projections.

\section{Discussion of some Applications}
\label{secDiscussion}

In order to illustrate the foregoing general results, we will in this
section discuss some applications to well known standard examples of
``realistic'' superfluid systems, ordered by increasing complexity.
\subsection{Single uncharged superfluid} 
Probably the simplest case 
are single, uncharged (isotropic) superfluids like $^4$He.
We note that vortices in $^3$He show a much richer structure than in
$^4$He (e.g.\ see \cite{LesHouches99}), due to the anisotropic type of
the microscopic Cooper pairing responsible for the superfluidity of
$^3$He. But the present approach should still be a good
approximation at least for the $^3$He--B superfluid \cite{volovik2000},
because sufficiently far from the vortex core the additional
(anisotropic) degrees of freedom of the order parameter are ``frozen''
and the dynamics is again mainly governed by the phase $\varphi^\sX$.
\\[0.2cm]
{\bf a) at $\bbox{T=0}$:} In the case of a single superfluid at zero
temperature, the ``entrainment matrix'' $K_{\X\Y}$ of
(\ref{equEntrainment}) reduces to \mbox{$K=n^0/\mu^0$}, where $n^\a$
is the particle current and $\mu_\a$ the momentum per particle of the
superfluid.  
There are no normal fluids, so $\S$ of (\ref{equMatrices}) is
given trivially by \mbox{$\S = K$}. 
The charge vector vanishes, \mbox{$e^\sX =0$}, 
and the charge projection tensors are trivial, so 
\mbox{$N^\sX_\ortho = N$} and \mbox{$N^\sX_\para =0$}. The vortex
energy (\ref{equResult1}) then simply reduces to
\be
U_\vort = N^2 \pi \hbar^2 {n^0 \over \mu^0} \ln {\ru\over \zeta}\,,
\fe
which is the same expression as found in \cite{cl95a} for the single
superfluid. In the nonrelativistic limit, where
\mbox{$\mu^0\rightarrow m$} and \mbox{$n^0 \rightarrow n$} (where $m$
is the rest mass of the superfluid particles, and $n$ their number
density), we recover
the usual expression for the (hydrodynamic) superfluid vortex energy
in the zero temperature limit (e.g.~see \cite{Tilley}).\\[0.2cm]
{\bf b) at $\bbox{T\not=0}$:} In the case of a finite temperature, the
system can be described as an effective superfluid--normal fluid
mixture, where the normal fluid consists of the viscous gas of
excitations in the superfluid. 
The superfluid and normal particle currents are $\cvec{n}_\sf$ and
$\cvec{n}_\nf$, and their respective momenta per particle
$\cvec{\mu}^\sf$ and $\cvec{\mu}^\nf$, say. There are no charged 
fluids, so $N^\sf_\para = 0$ and $N^\sf_\ortho = N$. The entrainment
matrix (\ref{equDefK}) reads 
\be
K_{\sX\sY} = 
\left( \begin{array}{cc}
K_{\sf\sf} & K_{\sf\nf} \\
K_{\nf\sf} & K_{\nf\nf} 
\end{array} \right) \,,
\fe
and is decomposed in the mongrel representation of
Sec.~\ref{secMongrel} as $\V=1/K_{\nf\nf}$, and 
$\S = K_{\sf\sf} - K_{\sf\nf}^2/K_{\nf\nf}$, so the vortex energy would
simply be given by inserting this expression for $\S$ into equation
(\ref{equResult1}).  However, in order to compare this result to the usual
expression for the vortex energy in superfluids at $T\not=0$, we have
to link the present entrainment formalism to the more common language
of Landau's two--fluid model \cite{Landau} that is expressed in terms
of a ``superfluid density'' $\rho_\sf$ and a ``normal density''
$\rho_\nf$. This ``translation'' has been achieved in a rigorous and
extensive manner by Carter and Khalatnikov \cite{ck94}, but for the
present purpose of an illustrative example, the following very simple
argument should show in a sufficiently convincing way how to translate
between the respective quantities.
Namely, consider the total (spatial) momentum density $T^{0i}$ (with
\mbox{$i=1,2,3$}) of the fluid mixture, for which from
(\ref{equStressEnergy}) we have 
\mbox{$p^i \equiv T^{0i} = n^0_\sf \mu^{\sf i} + n^0_\nf \mu^{\nf i}$}.
Using the mongrel relations (\ref{equLC1}) and (\ref{equLC2}), this can
be rewritten as 
\mbox{$p^i = (\mu^{\sf0} \S) \,\mu^{\sf i} + (n_\nf^0 \V)\, n_\nf^i$}.
Now we introduce the normal velocity 
\mbox{$v^i_\nf \equiv n^i_\nf / n_\nf^0 $}, which is the real
mean velocity of the excitations, and the superfluid
``pseudo--velocity'' 
\mbox{${\tilde v}^i_\sf \equiv \mu^{\sf i} / \mu^{\sf0}$}, which is
not a ``real'' velocity in the sense of a particle transport. 
In the nonrelativistic limit, where $\mu^{\sf0}$
tends to the constant rest mass of the superfluid particles, 
\mbox{$\mu^{\sf0}\rightarrow m_\sf$}, the
irrotationality property of superfluids (\ref{equIrrot}) implies
\mbox{$\nabla_{\rule{0em}{1ex}}^{[i}{\tilde v_\s}^{j]}\approx0$}, in
other words  \mbox{``$\text{rot}\,\vec{\tilde v}_\s = 0$''}.  
In these variables the total momentum density now reads
\be
p^i = \left[ (\mu^\sf_0)^2 \,\S \right] \tilde{v}_\sf^i + 
\left[ (n^0_\nf)^2 \,\V \right] v_\nf^i\,.
\fe
Comparing this to the orthodox expression 
\be
p^i = \rho_\sf \tilde{v}_\sf^i + \rho_\nf v_\nf^i\,,
\fe
we can identify 
\be
\rho_\sf =  (\mu^\sf_0)^2 \,\S \,,\quad\text{and}\quad
\rho_\nf = (n^0_\nf)^2 \,\V \,.
\label{equMatchLandau}
\fe
This is consistent with the additivity postulate 
\mbox{$\rho = \rho_\sf + \rho_\nf$}, namely using (\ref{equCross}) we
obtain the expression 
\mbox{$\rho = n_\sf^0 \mu^{\sf0} + n_\nf^0 \mu^{\nf0}$}, which effectively
reduces to the total mass density in the Newtonian limit.
In the present case we have \mbox{$\mu^{\sf0} \rightarrow m_\sf$} for
the superfluid, while \mbox{$\mu^{\nf0} \rightarrow 0$}, as the normal
fluid is identified with the gas of excitations, so 
the total mass density reduces to \mbox{$\rho \rightarrow n_\sf m_\sf$}.

In the nonrelativistic limit, expression (\ref{equMatchLandau}) yields 
\mbox{$\S = \rho_\sf / m_\sf^2$}, and so the equation (\ref{equResult1})
for the vortex energy can explicitly be written as
\be
U_\vort = \pi \hbar^2 N^2 {\rho_\sf \over m_\sf^2}\,\ln {\ru \over \zeta}\,,
\fe
in agreement with the well known result in Landau's two--fluid model
(e.g.\ see \cite{Tilley}).
\subsection{Two uncharged superfluids}
In the next step, let us consider a vortex in a mixture of two uncharged
superfluids, as first considered by Andreev and Bashkin \cite{ab75}
for a mixture of $^3$He and $^4$He. 
Again, at $T=0$ there are no normal fluids, so we have
\be
\S_{\sX\sY} = K_{\sX\sY} = 
\left( \begin{array}{cc}
K_{33} & K_{34} \\
K_{43} & K_{44} 
\end{array} \right) \,.
\fe
The charge vector vanishes, \mbox{$e^\sY =0$}, and so
\mbox{$N^\sX_\para = 0$} and \mbox{$N^\sX_\ortho=\{ \,N^3,\,N^4\,\}$}. 
The expression (\ref{equResult1}) for the vortex energy in this case
explicitly reads 
\bea
U_\vort = \pi\hbar^2 \big[ (N^3)^2 K_{33} &+& (N^4)^2 K_{44} \nonumber\\
& &+ 2 N^3 N^4 K_{34}\big] \ln {\ru\over \zeta}\,.
\fea
We see that there is a purely hydrodynamic interaction energy due to 
entrainment (i.e.\ not related to the condensation energy in the core)
from the last term in brackets, which is either attractive or 
repulsive depending on the sign of the entrainment coefficient
$K_{34}$.
\subsection{Conventional Superconductors}
When we consider cases with charged superfluids, the simplest
example is already a two constituent system, because a second charged
component is necessary to allow for global charge neutrality.
This picture applies for example to conventional laboratory
superconductors, where the charged superfluid (charge $e^-$ and 
particle density $n_-$) consists of Cooper paired conduction
electrons, while the second component is the ``normal'' background of
positively charged ions (charge $e^+$ and particle density $n_+$).
In the maximally symmetric and stationary situations considered in the
present work, ``normal'' components are naturally restricted to
uniform rotation (\ref{equNormalCurrent}), and therefore it makes
no difference whether the normal component is actually a
real ``fluid'' or a solid lattice like in the present example.

Because of the Cooper pairing mechanism, the fundamental superfluid
charge carriers have to be considered as electron pairs, and therefore
the charge per superfluid particle $e^-$ should be twice the electron
charge, i.e.\ \mbox{$e^-= -2e$}, and consequently the rest mass is
\mbox{$m^-=2m_\e$}, where $m_\e$ is the electron rest mass. 
The entrainment matrix $K_{\X\Y}$, defined in (\ref{equDefK}) can be
written as 
\be
K_{\X\Y} = \left( \begin{array}{cc}
K_{--} & K_{-+} \\
K_{+-} & K_{++}
\end{array} \right) \,,
\fe
and the transformation into the mongrel representation of
Sec.~\ref{secMongrel} yields  
\mbox{$\S = K_{--} - (K_{-+})^2/K_{++}$}
and \mbox{$\V = 1/K_{++}$}. 
The charge vector is just \mbox{$e^\sX = e^-$}, and so 
\mbox{$N^\sX_\ortho=0$} and \mbox{$N^\sX_\para=N$}.\\[0.2cm] 
{\bf The London field:}
In the simple case of a vortex--free state, i.e.\ with $N=0$, there is
nevertheless a nonvanishing uniform London field $B_\bg$ if the
superconductor is rotating (rotation rate $\Omega$).
Equation (\ref{equLondon2}) for the London field immediately yields
for this simple case \mbox{$B_\bg = -2\Omega ({E / e^-})$}, where $E$
is the energy per superfluid particle, i.e.\ \mbox{$E=-\mu^-_0$}.
If we choose a reference frame with \mbox{$L\equiv \mu^-_z = 0$},
i.e.\ comoving with the superconductor in $z$ direction, then $E$ can
be identified with the (relativistic) chemical potential 
\mbox{$\mu^- \equiv (-\mu^-_\a \mu^\a_-)^{1/2}$}. In the Newtonian
limit, where \mbox{$\mu^- \approx m^-$}, the conventional Newtonian
chemical potential $\mu^-_{\rm chem}$ is related to the relativistic
chemical potential $\mu^-$ as
\be
\mu^- = m^- \left( 1 + {\mu^-_{\rm chem}\over m^-} + \O( \epsilon^2 )\right)\,,
\fe
where \mbox{$\epsilon\equiv \mu^-_{\rm chem}/m^- \ll 1$}.
The London field for a rotating superconductor can therefore be
written in the form
\be
B_\bg = -2\Omega {m^-\over e^-}
\left(1 + {\mu^-_{\rm chem} \over m^-} + \O(\epsilon^2) \right) \,.
\label{equLondon99}
\fe
It is well known that that the ``entrainment'' formalism for
interacting constituents can equivalently be expressed in the more
conventional (albeit sometimes less convenient) language of
``effective masses''  \cite{ab75}. We see that in the case of
two--constituent superconductors, the effect of entrainment
(i.e.\ effective masses) cancels out in the expression
(\ref{equLondon99}) for the London field, which therefore depends quite
naturally on the ``bare'' electron rest mass to charge ratio
$m^-/e^-$, including a ``relativistic'' correction due to the finite
chemical potential $\mu^-_{\rm chem}$ of the electrons. 
We note that this cancellation only occurs for systems with a single
superfluid constituent, where $\S$ is consequently a scalar and
cancels out in (\ref{equLondon2}). As soon as there is a second
(interacting) superfluid constituent involved, as in the following
example of neutron star matter, the London field {\em does} depend on
the effective masses of the constituents.  
We further note that the present covariant treatment is intrinsically
frame--independent, and contrary to the analysis of \cite{liu98},
we find that $B_\bg$ does {\em not} depend on the chemical potential
$\mu^+$ of the ``normal'' component of positively charged ions. 

A very crude estimate of the relativistic correction term 
\mbox{$\mu^-_{\rm chem}/m^-$} for a Nb superconductor at $T=0$, taking 
$\mu^-_{\rm chem}$ simply to be the Fermi energy of a free electron gas,
yields a (positive) correction of the order $10^{-4}$. This is in
qualitative and nearly quantitative agreement with precision
measurements performed on a rotating Nb superconductor
\cite{cabrera89}. But in order to effectively compare expression
(\ref{equLondon99}) with experimental results, a more careful
estimation of $\mu^-_{\rm chem}$ would be necessary.\\[0.2cm]
{\bf Vortices:}
Now let us consider a  vortex configuration, i.e.\ with
\mbox{$N\not=0$}. We see that a similar cancellation of the entrainment 
effect as for the London field (\ref{equLondon99}) arises for the
total flux of the vortex, which is seen by (\ref{equFlux}) to give the
usual 
\be
\Phi = N \Phi_0\,,\quad\text{with}\quad 
\Phi_0\equiv {2\pi\hbar\over e^-} \,,
\fe
while the London penetration depth (\ref{equPenetration}) {\em is} modified
by entrainment, namely \mbox{$l^{-2} = 4\pi (e^-)^2 \S$}. To write this more
explicitly, we note that $\S$ can be written in the absence of
entrainment as \mbox{$\S^{(0)} = n_-/\mu^-$} and further 
\mbox{$\S^{(0)} = (n_-/m^-) (1 - \delta_{\rm rel})$},
where \mbox{$\delta_{\rm rel} \equiv \mu^-_{\rm chem}/m^-$} is the same
relativistic correction factor encountered in the expression for the
London field (\ref{equLondon99}). A nonvanishing entrainment
interaction between the constituents will add an additional correction
term $\delta_{\rm entr}$ proportional to the matrix element $K_{+-}$,
so that $\S$ can be written as 
\mbox{$\S = (n_-/m^-) (1 + \delta_{\rm entr} - \delta_{\rm rel})$},
and so the London penetration depth reads
\be
\l^{-2} = 4\pi (e^-)^2 {n_- \over m^-}
(1+\delta_{\rm entr} - \delta_{\rm rel}) \,.
\fe 
The vortex energy is given by the ``magnetic'' term in
(\ref{equResult1}) alone, due to \mbox{$N_\para = N$} and
\mbox{$N_\ortho =0$}, so we recover the usual ``axis--field''
expression 
\be
U_\vort \approx {\Phi \proper{B}(0) \over 8\pi} \,,
\fe
which is seen in the more explicit form (\ref{equResultII}) (for the
type--II limit, for simplicity) to depend on the effect of entrainment,
namely 
\be
U_\vort = N^2 \pi\hbar^2 {n_-\over m^-}(1+\delta_{\rm entr} -
\delta_{\rm rel}) \ln{\l\over\xi} \,,
\fe
but as the total vortex energy \mbox{$U = U_\vort + U_\con$} also depends on
the largely unknown condensation energy of the core, the relativistic
and entrainment corrections in this expression seem unlikely to be of
observable interest.
\subsection{Neutron star matter (Outer core)}
In this last example we consider the case of a (cold) degenerate
plasma consisting of neutrons, protons and electrons in $\beta$
equilibrium, as relevant for the outer core of neutron stars (i.e.\ at
densities $\gtrsim$ nuclear density). 
In this case one usually assumes that there is an important entrainment 
between neutrons and protons due to their strong interactions, while
the entrainment with electrons is generally supposed to be
negligible. We will follow this assumption and denote the
entrainment matrix as
\be
K_{\X\Y} = \left( \begin{array}{ccc}
K_{\n\n} & K_{\n\p} & 0 \\
K_{\p\n} & K_{\p\p} & 0 \\
 0	&  0	& K_{\e\e} \end{array} \right) \,.
\fe
The calculations of the superfluid gaps  for this neutron star matter 
generally suggest (see for example \cite{bcjl92}) that the protons 
will be superconducting and the neutrons superfluid, while the
electrons remain ``normal'', so this system would represent a
superconducting--superfluid--normal mixture.
The matrices of the mongrel representation of Sec.~\ref{secMongrel}
for this system read \mbox{$\M=0$}, \mbox{$\V = 1/K_{\e\e}$} and 
\be
\S_{\sX\sY} = \left( \begin{array}{cc}
K_{\n\n} & K_{\n\p} \\
K_{\p\n} & K_{\p\p} \end{array} \right) \,,
\fe
and we define an ``entrainment coefficient'' 
\mbox{$\alpha \equiv K_{\n\p}/K_{\p\p}$}.
For this system the charge vectors and projections are nontrivial,
namely 
\be
e^\sY = \left( \begin{array}{c}
0\\ q \end{array}\right) \,,\quad 
e_\sY = q K_{\p\p} \left( \alpha, \,1 \right) \,,
\fe
\be
\eta^\sX_\sY = \left( \begin{array}{cc}
0 & 0 \\
\alpha & 1 \end{array}\right)\,,\quad
\gamma^\sX_\sY = \left(\begin{array}{cc}
1 & 0 \\
-\alpha & 0\end{array}\right) \,,
\fe
where $q$ is the charge of a proton Cooper pair, i.e.\
\mbox{$q=2|e|$}, and we further have
\bea
N^\sX_\para &=& (N^\p + \alpha N^\n) \left(\begin{array}{c}
0 \\ 1 \end{array}\right) \,,\\
N^\sX_\ortho &=& N^\n \left(\begin{array}{c}
1 \\ -\alpha \end{array}\right)\,.
\fea
The London penetration depth (\ref{equPenetration}) is
\be
\l^{-2} = 4\pi q^2 K_{\p\p} \,,
\fe
and the vortex flux (\ref{equFlux}) is found as
\be
\Phi = (N^\p + \alpha N^\n) \Phi_0\,, \quad\text{with}\quad
\Phi_0 = {2\pi\hbar \over q} \,,
\fe
in agreement with earlier results in the literature
\cite{ss80,vs81,als84}. 
The vortex energy in the type--II limit (\ref{equResultII}) reads
\bea
U_\vort &= &\pi\hbar^2 (N^\n)^2 \left[ 
(K_{\n\n} - \alpha^2 K_{\p\p}) \ln {\ru\over \xi} + \alpha^2
K_{\p\p}\ln{\l\over\xi} \right] \nonumber\\
& & + \pi\hbar^2 (N^\p)^2 K_{\p\p} \ln{\l\over\xi}\label{equNPVortex}\\
& & + 2 \pi\hbar^2 N^\n N^\p K_{\n\p} \ln{\l\over\xi}\,.\nonumber
\fea
Similar to the case of a mixture of two uncharged superfluids, we see
that the total vortex energy consists of a pure n--vortex term,
a pure p--vortex term (each of which is
modified by the entrainment),  while the last term
represents an attractive or repulsive (depending on the sign of
$K_{\n\p}$) interaction term with respect to infinite separation.
It has been suggested \cite{sed95} that the effect of entrainment
between neutrons and protons could energetically favor a ``vortex
cluster'' structure (i.e.\ a neutron vortex surrounded by a dense
lattice of proton vortices) with respect to a single neutron vortex. 
This question can strictly speaking not be addressed in the
present framework of perfectly axially symmetric configurations, and
will be subject of future investigation, but the energy of a
single n--vortex ($N^\p=0$, $N^\n\not=0$) is seen from expression
(\ref{equNPVortex}) to be of the same order of magnitude if not
smaller than in the absence of entrainment ($\alpha\rightarrow 0$),
i.e. 
\mbox{$U^{(0)}_\vort = \pi \hbar^2 (N^\n)^2 K^{(0)}_{\n\n}\ln({\ru/\xi})$}.
Any configuration containing more vortices is therefore
rather expected to have a higher energy, but  the possibly
attractive interaction term in (\ref{equNPVortex}) could lead to an
effective ``clustering'' of already present vortices, namely a
n--vortex that ``accretes'' p--vortices until saturation.

\acknowledgments
The author thanks B.~Carter and D.~Langlois for many instructive
discussions and helpful advice.

\end{document}